\documentclass{mem}
\usepackage{natbib}
\usepackage{txfonts}
\usepackage{balance}
\usepackage{graphicx}
\usepackage[breaklinks,dvipdfm]{hyperref}
\idline{75}{282}

\begin{document}
\title{
SDSS J080434.20+510349.2: Cataclysmic Variable Witnessing  the
Instability Strip? }

   \subtitle{}

\author{
E. \,Pavlenko\inst{1}, V. \,Malanushenko\inst{2}, G.
\,Tovmassian\inst{3}, S. \,Zharikov\inst{3},  T. \,Kato\inst{4},\\
N. \,Katysheva\inst{5}, M. \,Andreev\inst{6}, A.
\,Baklanov\inst{1}, K. \,Antonyuk\inst{1}, N. \,Pit\inst{1}, A.
\,Sosnovskij\inst{1}, S. \,Shugarov\inst{5,7}
          }

  \offprints{E. Pavlenko}

\institute{ Crimean Astrophysical Observatory, Crimea 98409,
Ukraine \email{eppavlenko@gmail.com} \and Apache Point
Observatory, P.O. Box 59, Sunspot, NM88349, USA \and Institute of
Astronomy, UNAM, AP877, Ensenada, Baja California, 22800 Mexico
\and Dep. of Astronomy, Kyoto University, Sakyo-ku, Kyoto
606-8502, Japan \and Sternberg Astronomical Institute, Moscow
University, Universitetsky Ave., 13, Moscow 119992, Russia \and
Institute of Astronomy, Russian Academy of Sciences, 361605, Peak
Terskol, Kabardino-Balkaria, Russia \and
Astronomica Institute of the Slovakia Academy of Sciense, 05960, 
Tatranska Lomnica, the Slovak Republic\\
 }

\authorrunning{Pavlenko }

\titlerunning{SDSS J080434.20+510349.2}

\abstract{ SDSS J080434.20+510349.2 is the 13th dwarf nova
containing a pulsating white dwarf. Among the accreting pulsators
that have experienced a dwarf novae outburst, SDSS J0804 has the
most dramatic history of events within a short time scale: the
2006 outburst with 11 rebrightenings, series of December 2006 --
January 2007 mini-outbursts, the 2010 outburst with 6 rebrightenings. 
Over 2006--2011, SDSS J080434.20+510349.2 in addition to positive 
$0.060^{d}$ superhumps during the outburst and 1-month post-outburst stage, 
$0.059005^{d}$ orbital humps in quiescence, displayed a significant short-term 
variations with periods P1 = 12.6 min, P2 = 21.7 min, P3 = 14.1 min and 
P4 = 4.28 min. The 12.6-min periodicity first appeared 7 months
after the 2006 outburst and was the most prominent one during the
following $\sim 900$ days. It was identified as non-radial
pulsations of the white dwarf. The period of this pulsations
varied within a range of 36 s, and amplitude changed from
$0.013^{m}$ to $0.03^{m}$. Simultaneously one could observe the
21.7-min and 14.3-min periodicities  with a much lower significance level. 
During the minioutbursts the 21.7-min  periodicity became the most powerful, 
the 12.6-min periodicity was less  powerful, and the 12.6-min periodicity 
had the lowest significance. After  the 2011 outburst the most prominent 
short-term periodicity appeared  $\sim 7$ months after the outburst, but at 
4.28 min. We identified that variability with periods P2, P3 and P4 could be 
additional pulsation modes, however the relation of P4 to white dwarf
pulsation also can't be excluded.
 \keywords{Stars: binaries -- Stars:
cataclysmic variables -- dwarf nova, Individual: -- Stars: SDSS
J080434.20+510349.2 } } \maketitle{}

\section{Introduction}

SDSS J080434.20+510349.2 (hereafter SDSS J0804) was first
discovered as a CV and was considered as a potential dwarf nova
with an underlying white dwarf (WD) in quiescence prior to the
2006 outburst \citep{Szkody06} and it was first found in outburst
by \citep{pav07}. Its 0.060-d superhump period
\citep{pav07, kato11} and 0.059005-d orbital period
\citep{pav09a,Zhar08,kato09} suggested SDSS J0804 as a period
bouncer, passed the period minimum \citep{Zhar08}. \citet{pav07}
first discovered the non-radial pulsations of the WD that appeared
8 months after the 2006 outburst and lasted for $\sim 2$ years.
Now there are thirteen accreting pulsating WDs belonging to the SU
UMa stars \citep{muk10}. Among the accreting pulsators that have
experienced a dwarf novae outburst, SDSS J0804 has the most
dramatic history of events in a short time scale. During 2006 --
2010 interval SDSS J0804 underwent two outbursts in 2006 and 2010,
accompanied by 11 and 6 rebrightenings consequently
\citep{pav07,kato11}. Additionally a series of mini-outbursts have
been observed in December, 2006 -- January, 2007 \citep{Zhar08}.
This  CV gives a unique opportunity to study the evolution of the
WD pulsations under such rare conditions.

\section{Data set}

Here we present an analysis using already published data
\citep{pav07,Zhar08,pav09a,pav09,kato09,pav10,kato11} and
unfiltered data newly obtained on April 4, September 5 and
September 6 at the 2.6-m Shajn telescope of the Crimean
astrophysical observatory with 20-s exposure time.

\section{2006 and 2010 outbursts}

Despite there was some reason to believe that SDSS J0804 is
similar to WZ Sge, possessing the outburst activity once per tens
of years \citep{pav07}, it displayed the second outburst four
years after the 2006 outburst. The overall light curve is shown in
Fig.1.

\begin{figure}[]
\resizebox{\hsize}{!}{\includegraphics[clip=true]{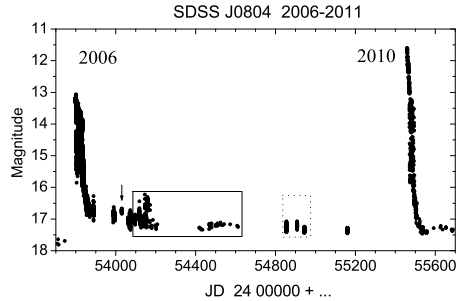}}
\caption{ \footnotesize The light curve of SDSS J0804 in
2006--2010. The arrow points to the first detection of pulsations.
The first prolonged box designates the time interval where
pulsations were observed. The second small box indicates the
period of unstable appearance of observed pulsations.}
\label{pavlenko_CVs_2011_01_fig01.eps}
\end{figure}

The increased scattering around JD = 2454150 is attributed to
minioutbursts. Note that the  first appearance of pulsations was
detected before and close to the start of minioutbursts.

The 2010 outburst differed from the previous one at least in a
sequence of rebrightenings. The comparison of rebrightenings is
given in Fig. 2. Besides their different number during the first
and second outbursts, rebrightenings in 2010 have a larger
amplitude. Also the system became fainter much more quickly in
2010 than in 2006 at the same epoch following the end of the main
outburst. In $\sim$ one month since the end of the main outburst,
SDSS J0804 was one magnitude fainter in the 2010 than in the 2006.

\begin{figure}[]
\resizebox{\hsize}{!}{\includegraphics[clip=true]{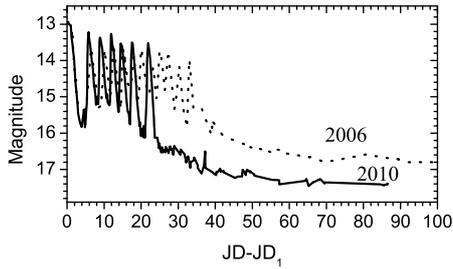}}
\caption{ \footnotesize The comparison of the rebrightenings
occured after the 2006 outburst (dotted line) and the 2010
outburst (solid line). The zero-point of the X-axis starts at the
rapid decline after the main outburst plateau. Data are combined,
using T0 = JD 2453801 for the 2006 outburst and T0 = JD 2455470
for the 2010 outburst.} \label{pavlenko_CVs_2011_01_fig02.eps}
\end{figure}

\section{WD pulsations}

During 50--60 days after both the 2006 and the 2010 outbursts the
most powerful signal of variation was the 0.060 d superhump period
with amplitudes of $0.15^{m} - 0.25^{m}$. Later it was replaced by
the orbital signal. The profile of the orbital light curve before
and after the 2010 outburst is similar. It was two-humped curve
with eclipses \citep {kato09} and with slight ($\sim 0.02^{m}$)
and occasional brightness depression in every hump. An example of
periodogram and data for 2011, April 30 folded on the orbital
period is shown in Fig. 3. The periodogram analysis has been
carried out using the Stellingwerf method \citep{pelt80}. Note
that there are no evidence of 12.6-min pulsations that have been
before the 2010 outburst. The period P3 roughly coincides with
Porb/6.  Period P2 will be discussed below.

\begin{figure}[]
\resizebox{\hsize}{!}{\includegraphics[clip=true]{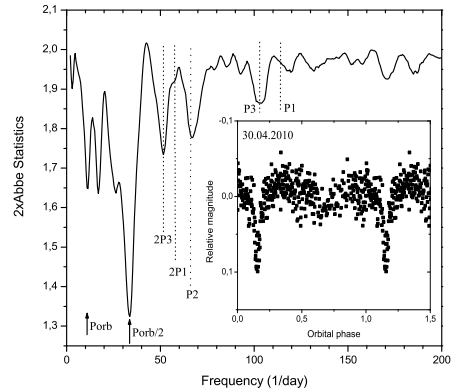}}
\caption{ \footnotesize Periodogram for the original data of April
30, 2011. In the inset the data folded on the orbital period are
shown. The peak corresponding to the 12.6-min pulsation is
designated as P1. The positions of periods corresponding to $P2 =
21.7$ min and $P3 = 14.1$ min (see later sections in the text for
detail) are also pointed.} \label{pavlenko_CVs_2011_01_fig03}
\end{figure}

The periodograms in a region of frequencies 10 -- 200 $d^{-1}$
during the 2007 mini-outbursts and apart of them differ
drastically (Fig. 4).

\begin{figure}[]
\resizebox{\hsize}{!}{\includegraphics[clip=true]{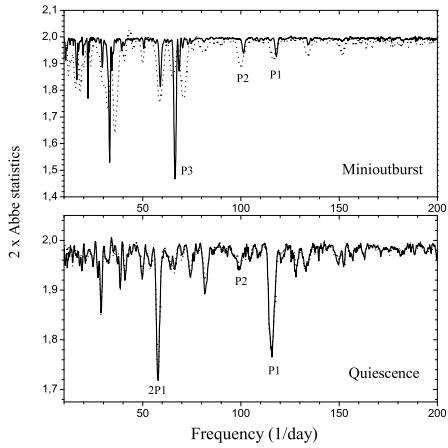}}
\caption{ \footnotesize Above: the two periodograms for the data
in the minioutburst (JD 2454117 and 2454118). Below: the two
periodograms for the data in quiescence (JD 2454123 and 2454124).
For each pair of periodogram the original ones  are denoted by the
solid and dotted lines. The orbital signal has been removed before
making the periodograms.} \label{pavlenko_CVs_2011_01_fig04}
\end{figure}

During the two nights in the minioutburst (JD 2454117 and 2454118)
the 12.6-min pulsations have been detected but at a much lower
 significance than the most prominent periodicity at 21.7
 min (designated as P2 in Fig. 3.) The 21.7-min period is not related to 
 harmonics of orbital period and we
 interpreted that it could be an independent  nonradial  pulsation of the 
 white dwarf. Meanwhile the periodograms for the data after  termination 
 of this minioutburst (JD 2454123 and 2454124) display the biggest peaks 
 related  to $P1 = 12.6$-min pulsations and 2P1. The peak related to P1 is 
 more  significant than those in the periodograms for the minioutburst.
 The 21.7-min peak is difficult to distinguish  from the  noise of
 periodograms in quiescence.

 On every nigh when the 12.6-min period was recorded, we
 estimated its current value.  The drift of this period in a
 region of 732 s -- 768 s is obvious. This region is $\sim 10$ times  wider than those
 found by Mukadam et al. (2010) for SDSS J161033.64-010223.3. Note that
 variations of the period presented by Mukadam et al. (2010) referred
 to a time scale $\sim 100$ times shorter then ours.
 In Fig. 5 we  presented the details of period drift around the minioutbursts.

\begin{figure}[]
\resizebox{\hsize}{!}{\includegraphics[clip=true]{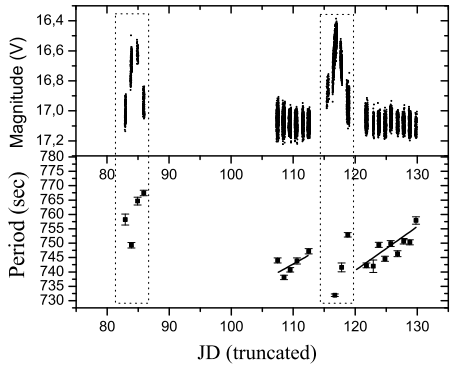}}
\caption{ \footnotesize The light curve including two
minioutbursts (above) an corresponding drift of 12.6-min pulsation
(below). The zero-point of the time scale is JD 2454000}
\label{pavlenko_CVs_2011_01_fig05}
\end{figure}

It is seen that during the minioutbursts themselves there is a
larger scatter of periods, while after the every of minioutburst
the periods lengthened. In a whole during the $\sim 900$ days this
period varied in the same region regardless of the presence of
minioutbursts. The drift of this period   together with amplitude
of pulsations is shown in Fig. 6. The amplitudes varied from
$0.013^{m}$ to $0.030^{m}$. The last data at JD 600 -- 960 showed
a decrease in period together with a decrease in amplitude. There
was, however, no correlation between periods and amplitudes for
the entire data.

\begin{figure}[]
\resizebox{\hsize}{!}{\includegraphics[clip=true]{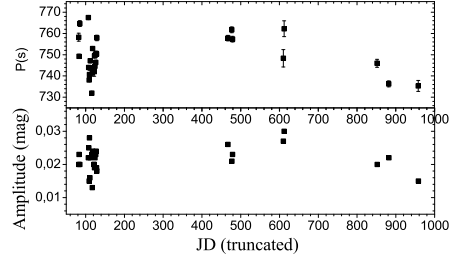}}
\caption{ \footnotesize The drift of periods and amplitudes of the
12.6-min pulsations of SDSS J0804 during $\sim 900$ days since the
appearance of pulsations. Zero-point of the time scale is  JD
245400.} \label{pavlenko_CVs_2011_01_fig06.eps}
\end{figure}

As already pointed out, the duration of the stable appearance of
12.6-min pulsations was $\sim 900$  days. After the 2010 outburst
these  pulsations did not appear at least during the following
$\sim 11$ months. However in the 7 months the 4.28-min (257 s)
periodicity was detected. We can't immediately claim that this
periodicity reflects a new pulsation mode. In Fig. 7 the
periodograms for the data of three nights (April 30, September 5
and 6, 2011) are shown.

\begin{figure}[]
\resizebox{\hsize}{!}{\includegraphics[clip=true]{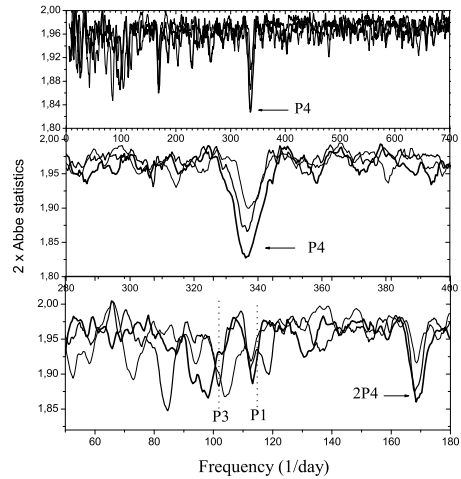}}
\caption{ \footnotesize Above: The periodograms for the data of
April 30 (thin line), September 5 (thick line) and September 6
(thickest line) in a region of frequencies 5--700 $d^{-1}$.
Middle: The central part of the periodograms. Below: The cut of
the periodograms in a region of peaks clustered at a long-period
part of periodograms. The position of periods P2 and P3 is shown
by dotted lines. The orbital signal was removed before making the
periodograms.} \label{pavlenko_CVs_2011_01_fig07.eps}
\end{figure}

One could see that a frequency, corresponding to the 4.28-min
period is the most prominent one in a region of frequencies of
5--700 $d^{-1}$ for all periodograms (there is also less
significant peak at 8.56-min that corresponds to its double
value). Its amplitude is $0.015^{m}$. It is impossible to conclude
immediately if this period is caused by the WD nonradial
pulsations or by the WD rotation (in the last case the expected
rotational period will be 8.56-min).

One could see also a group of a less significant peaks near the
long-period edge of the periodograms, clustering around frequency,
corresponded to period P3. From these three nights the periodogram
for the April, 30 and September, 5 indeed contained a peak which
coincides with P3 within the limit of  accuracy. Note that there
are no peaks at the 12.6-min period.

\section{Conclusions}

SDSS J0804 experienced  many accretion events over last four
years, that could affect the WD, resulting  in the appearance and
disappearance of WD nonradial pulsations.  In present work we
investigated the behavior of the most prominent
 12.6-min pulsations (P1), which was
stable over $\sim 900$ days. The nature of other periods is
unknown. Future observations are needed in order to clarify
whether or not the P2 = 21.7 min and P3 = 14.1 min have a
transient nature, these periods have relation to the WD
pulsations, and if the 4.28-min periodicity is a new pulsation
mode or related to the WD rotation.

\begin{acknowledgements}
E.P. and N.K. are grateful to Franco Giovannelli for his kind
invitation to participate in this Workshop. This work is partially
supported by the program "Kosmomikrofizika-2" of the Ukrainian
Academy of Science and grant NSh-7179.2010.2.
\end{acknowledgements}

\bibliographystyle{aa}

\begin{thebibliography}{}

\bibitem[Kato et al. (2009)]{kato09}
Kato, T.,~et al.\ 2009, \pasj, 61, 601

\bibitem[Kato et al. (2011)]{kato11}
Kato, T.,~et al.\ 2011, ArXiv
Astrophysics e-prints, 1108.5252

\bibitem[Mukadam et al. (2010)]{muk10}
Mukadam, A.~S.,~et al. \ 2010, \apj, 714, 1702

\bibitem[Pavlenko et al. (2007)]{pav07}
Pavlenko, E.~P.,~et al.\ 2007, ASP Conf. Ser., 372, 511

\bibitem[Pavlenko \& Malanushenko (2009)]{pav09}
Pavlenko, E. P.~\& Malanushenko, V. P.\ 2009, KPCB, 25, 48

\bibitem[Pavlenko (2009)]{pav09a}
Pavlenko, E. \ 2009, JPhCS, 172, 2071

\bibitem[Pavlenko et al. (2010)]{pav10}
Pavlenko, E., Antonyuk, O., Andreev, M., ~et al.\ 2010, AIP Conf.
Proc., 1273, 332

\bibitem[Pelt (1980)]{pelt80}
Pelt, Ya. \ 1980, Frequency analysis of astronomical time series,
Tallinn, Valgus Publ.

\bibitem[Szkody et al.(2006)]{Szkody06}
Szkody, P.,~et al.\ 2006, \aj, 131, 973

\bibitem[Zharikov et al. (2008)]{Zhar08}
Zharikov, S.~V.,~et al.\ 2008, \aap, 486, 505

\end{thebibliography}

\bigskip
\bigskip
\noindent {\bf DISCUSSION}

\bigskip
\noindent {\bf PAULA SZKODY:} It seems quite unusial to me 1-mag
minioutbursts in a WZ Sge star (they usually only have SOBs).

\bigskip
\noindent {\bf ELENA PAVLENKO:} SDSS J0804 indeed is unusual WZ
Sge type star. It has common features with WZ Sge stars such as a
short orbital period 0.0590048 d, large 6-mag amplitude of the
superoutburst, 11 rebrightenings after the 2006 outburst and 6
rebrightenings after the 2010 outburst, the two-humped orbital
profile. But occurrence of mini-outbursts is known only for this
system.

\bigskip
\noindent {\bf DMITRY BISIKALO:} Do you have any explanations of
these pulsations?

\bigskip
\noindent {\bf ELENA PAVLENKO:} Not yet! This question is rather
difficult. Probably these pulsations are more strong during the
mini-outburst and after them the amplitude of pulsations became
less then the amplitude of the 12.6-min pulsations.

\bigskip
\noindent {\bf DMITRY KONONOV:} What is the photometric precision
of the instrument you use and what mathematical methods do you use
to pick out these pulsation periods?

\bigskip
\noindent {\bf ELENA PAVLENKO:} All observations were done with
the 2.6-m Shajn telescope of the Crimean astrophysical observatory
in primary focus with best accuracy of the 0.005--0.007
magnitudes. The analysis we used was method of Stellingwerf
developed by Jaan Pelt  in his package "ISDA".

\end{document}